\documentclass[preprint,review,12pt]{elsarticle}

\usepackage{amssymb}

\usepackage{graphicx}      

\newcommand{\be}{\begin{equation}}
\newcommand{\ee}{\end{equation}}

\journal{Commun Nonlinear Sci Numer Simulat}

\begin{document}
\begin{frontmatter}

\title{Fractional Standard Map: Riemann-Liouville vs. Caputo}


\author[yu,nyu]{M.~Edelman\corref{cor1}}  
\ead{edelman@cims.nyu.edu}
\cortext[cor1]{Corresponding author}

\address[yu]{Department of Physics, Stern College at Yeshiva University, 
245 Lexington Ave, New York, NY 10016, USA}
\address[nyu]{Courant Institute of Mathematical Sciences, New York University,
251 Mercer St., New York, NY 10012, USA}


\begin{abstract}                
Properties of the phase space of the standard maps with memory obtained
from the differential equations with the Riemann-Liouville and Caputo 
derivatives are considered. Properties of the attractors which these
fractional dynamical systems demonstrate are different from properties of
the regular and chaotic attractors of systems without memory: they exist
in the asymptotic sense, different types of trajectories may lead to the
same attracting points, trajectories may intersect, and chaotic attractors
may overlap. Two maps have significant differences in the types of
attractors they demonstrate and convergence of trajectories to the
attracting points and trajectories.
Still existence of the the most remarkable new type of attractors, 
``cascade of bifurcation type trajectories'', is a common feature of both 
maps.  
\end{abstract}

\begin{keyword}
Discrete map \sep Fractional dynamical system \sep Attractor 
\sep Periodic trajectory \sep Map with memory \sep Stability  
\end{keyword}

\end{frontmatter}

\section{Introduction}
\label{int}

It is commonly accepted that fractional differential equations (FDE) play an
important role in the explanation of many physical phenomena. The physical 
systems that can be described by FDEs, physical fractional dynamical
systems (FDS), include Hamiltonian systems \cite{ZasBook:2005};  systems
of oscillators with long range interaction \cite{TZ3,TarZasEd};
dielectric \cite{JPCM2008-1} and viscoelastic \cite{MBook} materials; etc.

The FDSs with time fractional derivatives represent systems with
memory. Properties of such systems can be significantly different  from
the properties of the systems without memory. As in the case of the regular 
dynamical systems, the standard map (SM), or rather the
fractional standard map (FSM), is a good candidate to start
studies of the general properties of the FDSs. 
The first study of the map with memory 
derived from a differential equation, fractional standard map,  was
done in \cite{EdTar}. References to the prior research of the
one-dimensional maps with memory, which were not derived from the
differential equations, can be found in this article. In \cite{EdTar}
new types of attractors were found for the FSM derived from a
differential equation with the Riemann-Liouville 
fractional derivative (FSMRL) and
stability analysis was performed for the fixed and period two points. 

In this article we present the results of the study of the FSM derived
from a differential equation with the Caputo fractional derivative (FSMC) and
compare them with the new and previously \cite{EdTar} obtained 
results for the FSMRL. The results are based on a large, but not exhaustive,
number of simulations and the continuing investigation may reveal new 
properties of the FSMs. 

\section{Riemann-Liouville and Caputo Fractional Standard Maps}

\subsection{Equations and Initial Conditions}

The standard map in the form

\be 
p_{n+1} = p_n - K \sin x_n, \ \  x_{n+1} = x_n + p_{n+1}  \  \ ({\rm mod} \ 2\pi )
\label{SM}
\ee 
can be derived from the differential equation
\be
\ddot{x}+K \sin(x) \sum^{\infty}_{n=0} \delta \Bigl(\frac{t}{T}-(n+\varepsilon) \Bigr)=0,
\label{SMDE}
\ee
were $\varepsilon  \rightarrow 0+$.

The equations for Riemann-Liouville and Caputo  FSMs 
were obtained in \cite{TZ1} and \cite{T1}.
The Riemann-Liouville FSM can be derived from the differential equation with
the Riemann-Liouville fractional derivative describing a kicked system
\be \label{difFSMRL}
_0D^{\alpha}_t x+K\sin(x) \sum^{\infty}_{n=0} \delta \Bigl(\frac{t}{T}-(n+\varepsilon) \Bigr)=0, 
\quad (1 <\alpha \le 2) 
\ee
were $\varepsilon  \rightarrow 0+$, with the initial conditions 
\be
(_0D^{\alpha-1}_tx) (0+) = p_1, \   \
(_0D^{\alpha-2}_tx) (0+) = b  ,
\label{FSMRLic}
\ee
where
$$
_0D^{\alpha}_t x(t)=D^n_t \ _0I^{n-\alpha}_t x(t)=
$$
\be
\frac{1}{\Gamma(n-\alpha)} \frac{d^n}{dt^n} \int^{t}_0 
\frac{x(\tau) d \tau}{(t-\tau)^{\alpha-n+1}}  \quad (n-1 <\alpha \le n),
\label{RL}
\ee
$D^n_t=d^n/dt^n$, and $ _0I^{\alpha}_t$ is a fractional integral.

The Caputo  FSM can be derived from a similar equation with the Caputo
fractional derivative
\be \label{difFSMC}
_0^CD^{\alpha}_t x+K\sin(x) \sum^{\infty}_{n=0} \delta \Bigl(\frac{t}{T}-(n+\varepsilon) \Bigr)=0, 
\quad (1 <\alpha \le 2) 
\ee
were $\varepsilon  \rightarrow 0+$, with the initial conditions 
\be
p(0)=(_0^CD^{1}_tx) (0) = (D^1_tx) (0)=p_0, \   \  x(0)=x_0  ,
\label{FSMCic}
\ee
where
$$
_0^CD^{\alpha}_t x(t)=_0I^{n-\alpha}_t \ D^n_t x(t) =
$$
\be
\frac{1}{\Gamma(n-\alpha)}  \int^{t}_0 
\frac{ D^n_{\tau}x(\tau) d \tau}{(t-\tau)^{\alpha-n+1}}  \quad (n-1 <\alpha \le n),
\label{Cap}
\ee

After integration of equation (\ref{difFSMRL}) the 
FSMRL can be written in the form
\be \label{FSMRLp}
p_{n+1} = p_n - K \sin x_n ,
\ee

\be \label{FSMRLx}
x_{n+1} = \frac{1}{\Gamma (\alpha )} 
\sum_{i=0}^{n} p_{i+1}V^1_{\alpha}(n-i+1) 
, \ \ \ \ ({\rm mod} \ 2\pi ) ,
\ee
where 
\be \label{V1}
V^k_{\alpha}(m)=m^{\alpha -k}-(m-1)^{\alpha -k} 
\ee
and momentum $p(t)$ is defined as
\be \label{MomRL}
p(t)= \, _0D^{\alpha-1}_t x(t).
\ee 
Here it is assumed that $T=1$ and $1<\alpha\le2$. The condition $b=0$
is required in order to have solutions bounded at $t=0$ for
$\alpha<2$ \cite{EdTar}. 
In this form the FSMRL equations in the limiting
case $\alpha=2$ coincide with the equations for the standard map under the
condition  $x_0=0$. For consistency and in order to compare corresponding
results for all three maps (SM, FSMRL, and FSMC) all trajectories
considered in this article have the initial condition  $x_0=0$.

Integrating equation (\ref{difFSMC}) with the momentum defined as
$p=\dot{x}$ and assuming $T=1$ and $1<\alpha\le2$, 
one can derive the FSMC in the form
\be \label{FSMCp}
p_{n+1} = p_n 
-\frac{K}{\Gamma (\alpha -1 )} 
\Bigl[ \sum_{i=0}^{n-1} V^2_{\alpha}(n-i+1) \sin x_i 
+ \sin x_n \Bigr],\ \ ({\rm mod} \ 2\pi ), 
\ee
\be \label{FSMCx}
x_{n+1} = x_n + p_0 
-\frac{K}{\Gamma (\alpha)} 
\sum_{i=0}^{n} V^1_{\alpha}(n-i+1) \sin x_i,\ \ ({\rm mod} \ 2\pi ), 
\ee
It is important to note that the FSMC ((\ref{FSMCp}),   (\ref{FSMCx})) can
be considered on a torus( $x$ and $p$ mod  $2 \pi$), a cylinder 
($x$  mod  $2 \pi$),  or in an unbounded phase space,
whereas the FSMRL ((\ref{FSMRLp}), (\ref{FSMRLx}))  
can be considered only in a cylindrical or
an unbounded phase space. The  FSMRL  has no periodicity in $p$  
and cannot be considered on a torus. This fact is related to the
definition of the momentum (\ref{MomRL}) and initial conditions 
(\ref{FSMRLic}). The comparison of the phase portraits of  two FSMs
is still possible if we compare the values of the $x$ coordinates on
the trajectories corresponding to the same values of the maps' parameters.

\subsection{Stable Fixed Point}

The SM, the FSMRL, and the FSMC have the same fixed point at  $(0,0)$.
In the case of the SM this point is stable for $ K<K_{cr}=4$. In the case 
of the FSMRL the following system describes the evolution of 
trajectories near the fixed point 
\be \label{FixedRLp}
\delta p_{n+1} = \delta p_n - K \delta x_n ,
\ee
\be \label{FixedRLx}
\delta x_{n+1} = \frac{1}{\Gamma (\alpha )} 
\sum_{i=0}^{n} \delta p_{i+1}V_{\alpha}(n-i+1) .
\ee
The system describing the evolution of trajectories near fixed point  $(0,0)$
for the FSMC is
\be \label{FixedCp}
\delta p_{n+1} = \delta p_n -  \frac{K}{\Gamma (\alpha -1 )} 
\Bigl[ \sum_{i=0}^{n-1} V^2_{\alpha}(n-i+1) \delta x_i + \delta x_n \Bigr], 
\ee
\be \label{FixedCx}
\delta x_{n+1} = \delta x_{n} + \delta p_{0} -
\frac{K}{\Gamma (\alpha )} 
\sum_{i=0}^{n}V^1_{\alpha}(n-i+1) \delta x_i .
\ee
Direct computations using equations (\ref{FSMCp}), (\ref{FSMCx})
show that the critical curve $K_{cr}(\alpha)$ (see Fig.~1a) such that 
fixed point  $(0,0)$ is stable for $K<K_{cr}$ and unstable for  $K>K_c$
in the case of the FSMC is the same as the critical curve obtained from the 
semi-analytic stability analysis of the FSMRL in 
\cite{EdTar}. This curve can be described by the equation 
\be \label{Stable}
\frac{V_{\alpha l}K_{cr}}{2 \Gamma (\alpha )} = 1 ,
\ee
where
\be \label{Valphal} 
 V_{\alpha l}  =  \sum_{i=1}^{\infty} (-1)^{i+1} V^1_{\alpha}(i).
\ee
Taking into account that the equations of the fractional maps 
(\ref{FSMRLp}), (\ref{FSMRLx}) and  (\ref{FSMCp}), (\ref{FSMCx})
and the stability problems  (\ref{FixedRLp})-(\ref{FixedCx})  
for the maps contain convolutions, it is
reasonable to introduce the generating functions
\be \label{GenXP}
\tilde{X}(t)=\sum_{i=0}^{\infty }\delta x_i t^i \   \  {\rm and} \   \ 
\tilde{P}(t)=\sum_{i=0}^{\infty }\delta p_i t^i.
\ee
After the introduction 
\be \label{GenW1}
\tilde{W}^1_{\alpha}(t)= \frac{K}{\Gamma (\alpha)}
\sum_{i=0}^{\infty }[(i+1)^{\alpha-1}-i^{\alpha-1}]t^i,
\ee
\be \label{GenW2}
\tilde{W}^2_{\alpha}(t)=  \frac{K}{\Gamma (\alpha-1)}
 \Bigl(
1+\sum_{i=1}^{\infty }[(i+1)^{\alpha-2}-i^{\alpha-2}]t^i \Bigr) ,
\ee
the stability analysis can be reduced to the analysis 
of the asymptotic behavior at $t=0$ of the derivatives of  
the analytic functions
\be \label{GenXRL}
\tilde{X}(t)=\frac{p_0 \tilde{W}^1_{\alpha}(t)}{K}  
\frac{t}{1 - t \Bigl(1- \tilde{W}^1_{\alpha}(t) \Bigr)  }   
\ee
\be \label{GenPRL}
\tilde{P}(t)=p_0 \frac{1+  \tilde{W}^1_{\alpha}(t) }
{ 1-  t \Bigl( 1- \tilde{W}^1_{\alpha}(t) \Bigr) } 
\ee 
for the FSMRL and 
\be \label{GenXC}
\tilde{X}(t)=\frac{t p_0 + (1-t) x_0}  
{(1 - t) \Bigl(1- t +t\tilde{W}^1_{\alpha}(t) \Bigr)  }   
\ee
\be \label{GenPC}
\tilde{P}(t)= \frac{1}{1-t} \Bigl[ p_0 - t
\frac{t p_0 + (1-t) x_0}  
{(1 - t) \Bigl( 1- t +t\tilde{W}^1_{\alpha}(t) \Bigr)  }
\tilde{W}^2_{\alpha}(t)
 \Bigr]
\ee 
for the FSMC.

\subsection{Phase Space for $K<K_{cr}$ and $1< \alpha <2$}

In what follows, almost all results  are conjectures. They were obtained
by numerical simulations for some values of parameters $K$ and $\alpha$
and then verified for some additional values from the corresponding range 
of the parameters' values.

For the area preserving standard map the  stable fixed and periodic points
are elliptic points - zero Lyapunov exponent. They are surrounded by the
islands of regular motion. In the case of fractional maps the islands turn 
into basins of attraction associated with the points of attraction or 
slowly diverging attracting trajectories which evolve from the periodic
points as $\alpha$ decreases from two. For $K<K_{cr}$ and 
$1< \alpha <2$ we found no chaotic or regular trajectories. Two initially 
close trajectories that start in the area between   
basins of attraction at first diverge,
but then converge to the same or different attractors.

There are significant differences not only between properties of the
regular and fractional standard maps but also between phase space
structures of the FSMRL and FSMC. 
There is more than one way to approach an attracting fixed point of
the FSMRL. In Fig.~1b there are two trajectories for the FSMRL with
$K=2$ and $\alpha=1.4$. The bottom one is a fast converging trajectory
that starts in the basin of attraction, in which 
$x_n \sim n^{-1-\alpha}$  and  $p_n \sim n^{-\alpha}$ (see Fig.~1c). 
The upper trajectory is an example of the  attracting slow converging
trajectory (ASCT) introduced in \cite{EdTar} in which  
$x_n \sim n^{-\alpha}$  and  $p_n \sim n^{1-\alpha}$  (see Fig.~1d).
The trajectories that converge to the fixed point which start outside 
of the basin of attraction are ASCTs. In the case of the FSMC all
trajectories converging to the fixed point have the same asymptotic
behavior:  $x_n \sim n^{1-\alpha}$  and  $p_n \sim n^{1-\alpha}$  
(see Fig.~1e).

In both the FSMRL and FSMC considered on a cylinder the stable fixed point 
$(0,0)$ is surrounded
by a finite basin of attraction, whose width $W$ depends on the values of $K$
and $\alpha$. For example, for $K=3$ and   $\alpha=1.9$ the width of the 
basin of attraction is $1.6<W<1.7$ for the FSMRL (\cite{EdTar}). 
For the FSMC with the same parameters the width is   $1.7<W<1.8$. 
Simulations of thousands of trajectories with $p_0 < 1.6$ performed by the 
author, of which only 50 (with $1.6 \le p_0 < 1.7$) are presented in Fig. 1f, 
show only converging trajectories, whereas among 50 trajectories with
$1.7 \le p_0 <1.8$ in Fig~2b there are trajectories 
converging to the fixed point  $(0, 0)$ as well as some trajectories 
converging to the fixed points $(0, 2 \pi)$, $(0, -2 \pi)$, and $(0, -4 \pi)$.

\begin{figure}
\centering
\rotatebox{0}{\includegraphics[width=7.7 cm]{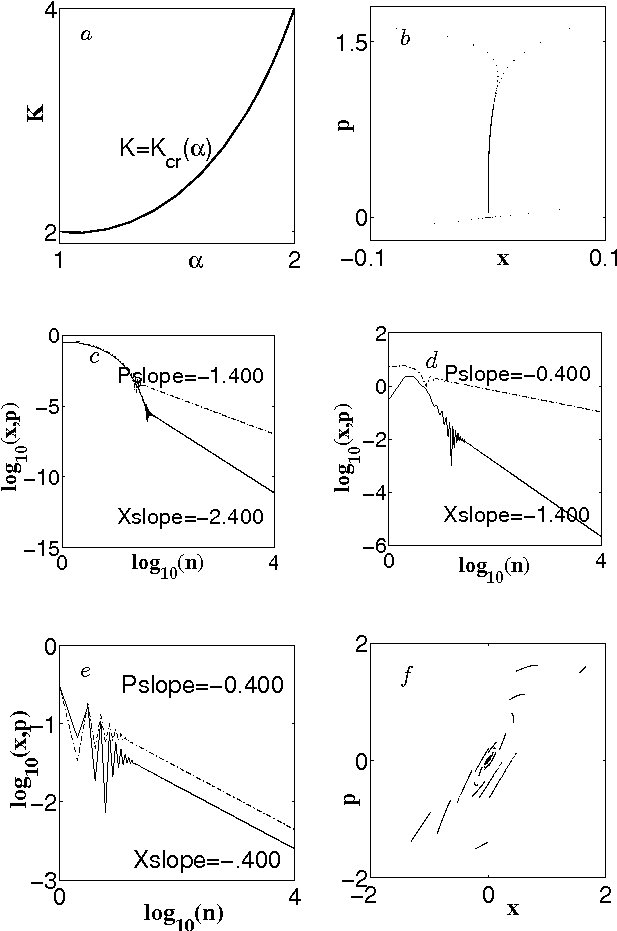}}
\caption{\label{Fig1}  The FSMRL and FSMC stable fixed point $(0,0)$: 
a). The fixed point $(0,0)$ for both the FSMRL and FSMC  is stable 
below the curve $K=K_{cr}(\alpha )$; 
b). Two trajectories for the FSMRL with $K=2$,  $\alpha =1.4$, and 
$10^5$ iterations on each trajectory. The bottom one with $p_0=0.3$ is a fast
converging trajectory. The upper trajectory with $p_0=5.3$ is an example
of the FSMRL's ASCT. The value of momentum on this ASCT 
after $10^5$ iterations is $p  \approx 0.042$;  
c). Time dependence of the coordinate and momentum for the fast converging 
trajectory from  Fig.~1b; 
d). $x$ and $p$ time dependence for the ASCT from the Fig.~1b; 
e). $x$ and $p$ time dependence for the FSMC with $K=2$, $\alpha =1.4$,
and $p_0=0.3$;  
f). Evolution of the FSMC trajectories with 
$p_0=1.6+0.002i$, $0 \le i < 50$ for 
the case $K=3$, $\alpha=1.9$. 
The line segments correspond to the $n$th iteration on the
set of trajectories with close initial conditions. 
The evolution of the trajectories with smaller $p_0$ and for the FSMRL
with  the same $K$, $\alpha$, and $p_0 < 1.6$ is similar.  
}
\end{figure}

\begin{figure}
\centering
\rotatebox{0}{\includegraphics[width=7.7 cm]{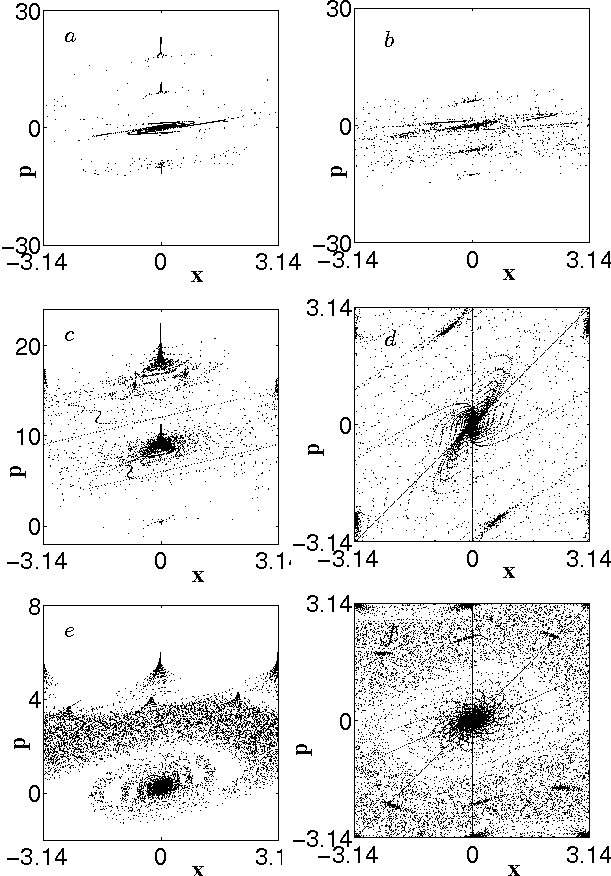}}
\caption{\label{Fig2} The  FSMRL and FSMC phase space for $K<K{cr}$: 
a). The FSMRL with the same values of parameters  as in
Fig~1f  but  $p_0=1.6+0.002i$, $0 \le i < 50$;
b). The FSMC with the same values of parameters  as in
Fig~1f  but  $p_0=1.7+0.002i$, $0 \le i < 50$;
c). 400 iterations on the FSMRL trajectories with 
$p_0=4+0.08i$, $0 \le i < 125$ 
for the case  $K=2$, $\alpha=1.9$. 
Trajectories converging to the fixed point, ASDTs with $x=0$, and
period 4 ASDTs are present;
d). 100 iterations on the FSMC trajectories with $p_0=-3.14+0.0314i$, 
$0 \le i < 200$ for the same case as in Fig.~2c ($K=2$, $\alpha=1.9$)
but considered on a torus. 
In this case all trajectories converge to the fixed point or period
four stable attracting points; 
e). 400 iterations on trajectories with $p_0=2+0.04i$, $0 \le i < 50$ 
for the FSMRL case $K=0.6$, $\alpha=1.9$. 
Trajectories converging to the fixed point and ASDTs of period 
2 and 3 are present; 
f).  100 iterations on the FSMC trajectories with $p_0=-3.14+0.0314i$, 
$0 \le i < 200$ for the same case as in Fig.~2e ($K=0.6$, $\alpha=1.9$)
considered on a torus. 
In this case all trajectories converge to the fixed point, period two and 
period three stable attracting points. 
}
\end{figure}

Another significant difference between two fractional standard maps is 
that all periodic points of the SM, except the fixed point $(0,0)$,
in the case of the FSMRL evolve into periodic attracting slowly diverging
trajectories (ASDT) (see Figs.~1a,~1c,~1e), whereas in the case of the
FSMC they evolve into the corresponding attracting points (see
Figs.~1b,~1d,~1f; in Figs.~1d,~1f the FSMC is considered on a torus).
Presence of only period one structures in the phase spaces 
of two fractional standard maps 
in the case $K=3$, $\alpha=1.9$  (Figs.~1a,~1b) corresponds to the fact
that the SM with  $K=3$ has only one central island.
Period $T=4$ structures in Figs.~1c,~1d and period  $T=2$
and $T=3$ in Figs.~1e,~1f correspond to the phase spaces of the SM with
$K=2$ and $K=0.6$.
Numerical evaluation (\cite{EdTar}) shows that ASDTs which converge to 
trajectories along the $p$-axis ($x \rightarrow x_{lim}=0$) have 
the following  asymptotic
behavior:  $x_n \sim n^{1-\alpha}$  and  $p_n \sim n^{2-\alpha}$.

\subsection{The FSMRL's Phase Space for $K>K_{cr}$ ($1< \alpha <2$)}

For  $K_{cr} < K < K_{cr2}(\alpha) \le 2 \pi$, the  FSMRL has period $T=2$ 
symmetric with respect to the origin stable
points with the  property
\be \label{T2point} 
p_{n+1} = -p_n, \    \  x_{n+1} = -x_n,
\ee
which evolve from  $T=2$  points  with the same property of the  SM, stable for
$4<K<2 \pi$. $K_{cr2}$ is the
upper with respect to $K$ limit of stability of these $T=2$ points. 

Assumption $(x_{n+1},p_{n+1})=(-x_n,-p_n)=(x_l,p_l)$ in 
system  ((\ref{FSMRLp}), (\ref{FSMRLx})) leads asymptotically 
($n\rightarrow \infty$) to 
\be \label{T2xl} 
x_l = \frac{K}{2 \Gamma(\alpha)} V_{\alpha l} \sin(x_l),
\ee
\be \label{T2pl} 
p_l = \frac{K}{2} \sin(x_l),
\ee
which has been analyzed in \cite{EdTar}. The condition of solvability
of (\ref{T2xl}) is
\be \label{T2solv} 
K > K_{cr}(\alpha) = \frac{2 \Gamma(\alpha)}{V_{\alpha l}},
\ee
which means that the stable $T=2$ points appear when the fixed point becomes
unstable. 

An example with $K=4.5$ of the evolution of the FSMRL's phase space
with the decrease
of $\alpha$ (from $\alpha$ on the critical curve or from 2 for $4<K<2\pi$) 
is presented in Fig.~3.  For large values of $\alpha$ ($1.74 < \alpha <2$ for
$K=4.5$) the phase space has a pair of stable symmetric $T=2$ attracting 
points (Fig.~3a), which in the case $4<K<2\pi$ evolve from the centers of 
the corresponding period two islands of stability of the SM. As in the case of
the fixed point, there are two types of convergence to the attracting points:  
slow with $\delta x_n \sim n^{-\alpha}$,  
$\delta p_n \sim n^{1-\alpha}$ and fast with
$\delta x_n \sim n^{-1-\alpha}$,  
$\delta p_n \sim n^{-\alpha}$. For lower values of $\alpha$
($1.67 < \alpha <1.73$ for $K=4.5$) there appears a couple of the 
non-symmetric stable
period two sets of attracting points (Fig.~3b), which with the decrease in 
$\alpha$  ($1.63 < \alpha <1.66$ for $K=4.5$) transform into attracting 
cascade of
bifurcations type trajectories (CBTT) (Figs.~3c,~3d). With the further decrease
in $\alpha$ the whole phase space of the FSMRL becomes chaotic (Fig.~3e).
When  $\alpha$ is close to one there appears a single chaotic attractor, which
at the lowest values of $\alpha$ ($\alpha <1.02$ for $K=4.5$) turns into
a set of disjoint chaotic attractors (Fig.~3f).

\begin{figure}
\centering
\rotatebox{0}{\includegraphics[width=9.2 cm]{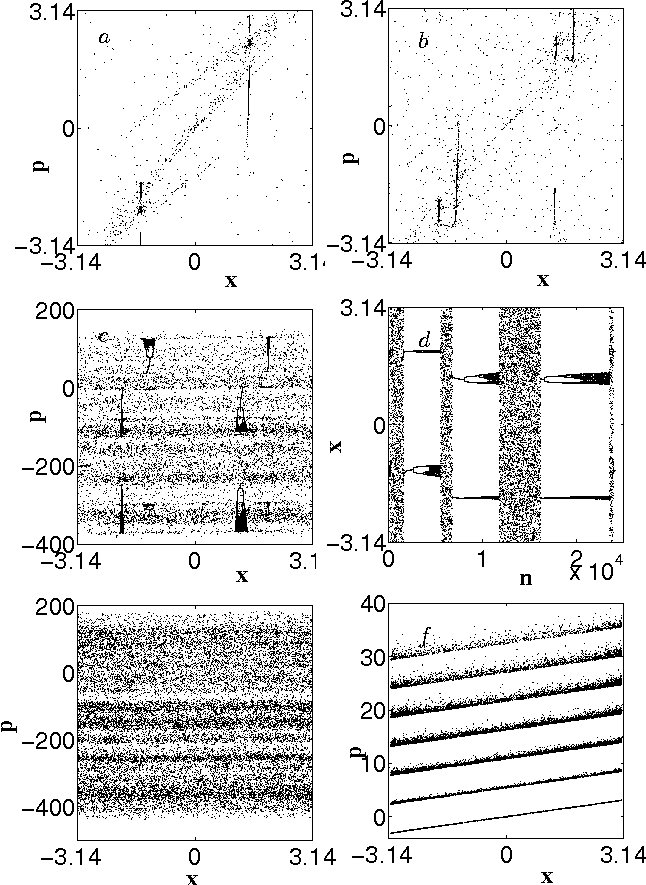}}
\caption{\label{Fig3}  The FSMRL's phase space for $K=4.5>K{cr}$: 
a). Period $T=2$ stable attracting points 
$x_{n+1}=-x_n$,  $p_{n+1}=-p_n$ 
for $\alpha=1.8$  500 iterations on each of 25 trajectories: 
$p_0=0.0001+0.08i$, $0 \le i <25$. Slow and fast converging trajectories.
b). Two sets of $T=2$ stable attracting points $x_{n+1} \ne -x_n$,  
$p_{n+1} \ne -p_n$ for $\alpha=1.71$. 500 iterations on trajectories with
the same initial conditions as in Fig.~3a;
c). 25000 iterations on a single trajectory with $\alpha=1.65$, $p_0=0.3$.
The trajectory occasionally sticks to one of the cascade of bifurcation 
type trajectories but then always recovers into the chaotic sea;
d). Time dependence of the coordinate $x$ in Fig.~3c; 
e). 20000 iterations on a single chaotic trajectory with  
$\alpha=1.45$, $p_0=0.3$; 
f). 7 disjoint chaotic attractors for $\alpha=1.02$. 1000 iterations 
on each of 20 trajectories: $p_0=0.0001+1.65i$, $0 \le i <20$. 
}
\end{figure}

The evolution of the FSMRM's phase space can also be considered for a fixed
value of $\alpha$  (in this paragraph we consider an example with 
$ \alpha = 1.6 $) and $K$ 
increasing from the value $K_{cr}$. 
When the value of $K$ is slightly above the critical curve the phase space has
a  pair of stable symmetric $T=2$ attracting points  
($K_{cr} < K <3.9$ for $\alpha=1.6$).
A couple of non-symmetric stable $T=2$ sets of attracting points
appears for  $4 < K <4.2$; CBTTs exist for  $4.3 < K <4.4$; 
for $K>4.5$ the whole phase space seems chaotic.

It should be noted that the existence of the  CBTTs can't be attributed to
the fact of the unusual definition of momentum in the derivation of the
FSMRL. The values of $x$-coordinates  in Fig.~3d represent the
solutions of the original differential equation (\ref{difFSMRL})
independently of the definition of momentum.

\subsection{The FSMC's Phase Space for $K>K_{cr}$ and $1< \alpha <2$}

\begin{figure}
\centering
\rotatebox{0}{\includegraphics[width=9.5 cm]{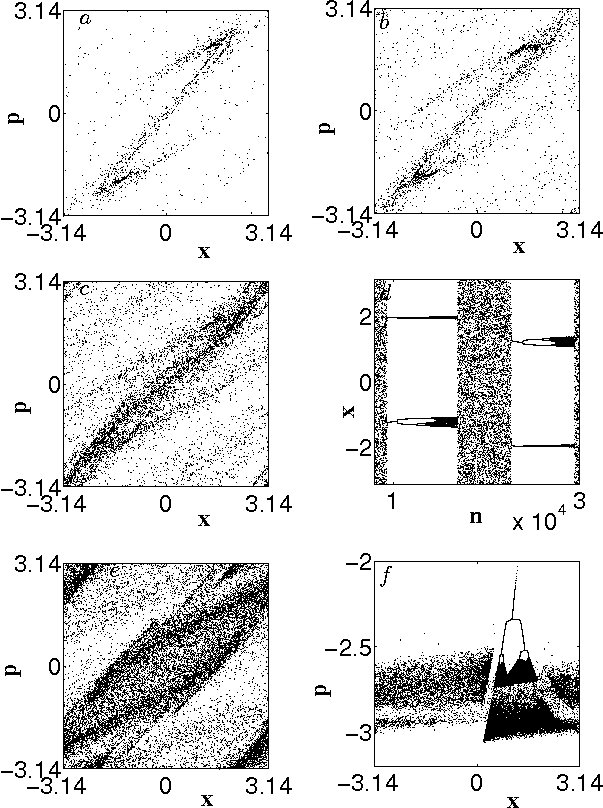}}
\caption{\label{Fig4}  The FSMC's phase space for $K=4.5>K{cr}$: 
a). Period $T=2$ stable attracting points 
$x_{n+1}=-x_n$,  $p_{n+1}=-p_n$ 
for $\alpha=1.8$. 1000 iterations on each of 10 trajectories: 
$p_0=-3.1415+0.628i$, $0 \le i <10$. 
b). Two sets of $T=2$ stable attracting points $x_{n+1} \ne -x_n$,  
$p_{n+1} \ne -p_n$ for $\alpha=1.71$. 1000 iterations on trajectories with
the same initial conditions as in Fig.~4a;
c). 30000 iterations on a single trajectory with $\alpha=1.65$, $p_0=0.3$.
The CBTTs can hardly be recognized on the full phase portrait, but can be seen
on the $x$ of $n$ dependence in Fig.~4d; 
d). Time dependence of the coordinate $x$ in Fig.~4c; 
e). 20000 iterations on a single trajectory with  
$\alpha=1.45$, $p_0=0.3$; 
f). 20000 iterations on each of two overlapping independent attractors 
for $\alpha=1.02$. The CBTT has $p_0=-1.8855$ and the chaotic attractor
$p_0=-2.5135$. 
}
\end{figure}
Numerical simulations show that the FSMC also has the period $T=2$ 
symmetric stable point for $K>K_{cr}$. After $10^5$ iterations the
difference in the values of $x_l$ for two maps is less than  $10^{-5}$.
The order of the difference in the values of $p_l$ is ten percent.
Fig.~4 gives an example of the evolution of the FSMC's phase space
with the decrease in $\alpha$ for $K=4.5$. The intervals of the existence 
of the symmetric $T=2$ points, asymmetric sets of the $T=2$ points,
and CBTTs are approximately the same for both maps. Trajectories 
converging to the symmetric $T=2$ points follow the same  law 
as the trajectories converging to the fixed point: 
$\delta x_n \sim n^{1-\alpha}$,  $\delta p_n \sim n^{1-\alpha}$.
It is difficult to resolve the CBTTs  on the phase portrait of the system
(Fig.~4c) and to recognize on a zoom but they are very clear on the $x$
of $n$ dependence in Fig~4d. The values of the $x$-coordinates of the CBTTs for
the FSMRL and FSMC are very close. For $\alpha <1.6$ the phase space
becomes chaotic with very nonuniform density of points (Fig.~4e). 
With the further decrease in $\alpha$ there appear chaotic attractors and then 
a set of disjoint and overlapping attractors. In Fig.~4f one can see two 
overlapping attractors for $\alpha = 1.02$. The CBTT on this picture 
(a dark set) persists even after 300000 iterations and overlaps with the
chaotic attractor. 

The observed overlapping of attractors and intersection of trajectories in
phase spaces is a consequence of the fact that the considered fractional
systems are systems with memory. In such systems the coordinates of the next
trajectory point are functions of the coordinates of all previously
visited points on a trajectory. In this case a coincidence of two points
does not lead to a coincidence of trajectories, and this property is  very 
different from the properties of phase spaces of the regular dynamical systems.

\section{Conclusion}
     
In this paper we continued the study of fractional maps started in 
\cite{EdTar}. We concentrated on the comparison of two maps 
at the values of the FSM parameter $K<2\pi$. Even though the simulations 
were not exhaustive, we were able to show that there are significant
similarities as well as differences between the structures of the phase
space of two FSMs. One of the findings is that the CBTTs are more
common in fractional maps than we originally thought. We hope that computer 
simulations of the equations of the physical systems that are described by 
the fractional differential equations will also produce the CBTTs and 
their origin will have a proper physical interpretation.

\section*{Acknowledgements}
The author expresses his gratitude to V.E. Tarasov and H. Weitzner 
for many comments and helpful discussions.
The author thanks DOE Grant 
DE-FG0286ER53223, and the Office of Naval Research Grant
No. N00014-08-1-0121 for the financial support.

\end{document}